# Classification of Power Quality Events in the Transmission Grid: Comparative Evaluation of Different Machine Learning Models

umut.guvengir@tubitak.gov.tr

**UMUT GÜVENGİR\*, DILEK KÜÇÜK, SERKAN BUHAN,**
*TÜBİTAK Marmara Research Center*

**CUMA ALİ MANTAŞ, MURATHAN YENİCELİ**
*Turkish Electricity Transmission Corp. (TEİAŞ)*

Turkiye

## SUMMARY

Automatic classification of electric power quality events with respect to their root causes is critical for electrical grid management. In this paper, we present comparative evaluation results of an extensive set of machine learning models for the classification of power quality events, based on their root causes. After extensive experiments using different machine learning libraries, it is observed that the best performing learning models turn out to be Cubic SVM and XGBoost. During error analysis, it is observed that the main source of performance degradation for both models is the classification of ABC faults as ABCG faults, or vice versa. Ultimately, the models achieving the best results will be integrated into the event classification module of a large-scale power quality and grid monitoring system for the Turkish electricity transmission system.

## 1. INTRODUCTION

Automatic detection, monitoring, analysis, and classification of power quality events are crucial for effective management of the electricity transmission grid. Classification of power quality events has been extensively studied in the literature, especially in the last two decades. In the context of designing the event classification module, relevant articles in the literature have been systematically reviewed and categorized in terms of the methods and algorithms used in these studies. Although obtaining raw data, dividing the data into parts, and processing the data using signal processing methods are common and standard stages in event classification studies, extracting the necessary features of the data to be used in classification algorithms, selecting the best features among these extracted features, and classifying these selected features using different methods and algorithms can increase the accuracy of classification. Various signal transformations are used in the feature extraction stage while metaheuristic optimization algorithms are preferred in the feature selection stage. Finally, machine learning algorithms, including deep learning-based algorithms, are used in the classification stage.

In this work, we present he evaluation results of different machine learning models for the classification of power quality events with respect to their root causes. At the end of our tests, the classification of possible root causes of events is achieved with high accuracy by using different machine learning algorithms. The best performing models will be integrated into the event classification module of a large-scale AI-based power quality and grid monitoring system designed and implemented for the Turkish electrical grid. Further comparative tests will also be conducted using deep learning models, as part of future work. In the rest of the paper, first a survey of relevant studies is presented, next, evaluation results of the machine learning-based event classification are provided, after a discussion of these significant results, the paper is concluded with a summary of points.

## 2. LITERATURE SURVEY

Articles published on event classification in the literature generally focus on the classification of event types itself, and there are relatively fewer studies on the classification of possible root causes of events. Since the methods and algorithms used in feature extraction, feature selection, and classification stages in event classifications can also be used in the classification of event causes, articles that do not focus on event causes have also been examined.

In the process of classifying the causes of an event, it is crucial to extract the necessary features from the signal by processing the raw data correctly. Classification algorithms only work efficiently when the relevant features of the signals are extracted. For this stage, many different signal processing methods have been proposed and employed in the literature. Wavelet Transform (WT) based methods have gained more popularity recently with the utilization of Continuous Wavelet Transform (CWT), and more frequently Discrete Wavelet Transform (DWT). Although the selection of the mother wavelet function is crucial for this method, the Daubechies-4 (db4) mother wavelet is mostly used in the literature as it detects fast transient signals with higher accuracy in power quality studies [1-3]. The DWT method is preferred more due to its minimum processing time and high accuracy compared to other signal processing methods which include Stockwell Transform (S-Transform), Hilbert Huang Transform (HHT), and Curvelet Transform (CT) [4-6]. After applying the selected signal processing method to the raw event data, statistical features such as mean value, maximum value, variance, standard deviation, RMS, kurtosis, skewness, and entropy are calculated to generate the scalar features of the signals to be utilized during the classification stage [3]. It is also commonly observed in the literature that metaheuristic optimization algorithms, clustering, and dimension reduction methods are utilized for selecting the optimal subset of the feature set to prevent high variance and overfitting problems in the classification stage [3, 7, 8].

Rule-based algorithms and expert systems were mainly used for classification purposes in the studies conducted in the last decade; however, machine learning algorithms, including deep learning-based algorithms, are applied more frequently in the current studies in parallel with the recent developments in the field of artificial intelligence [9]. Furthermore, it has been observed that classification of the root causes of events is performed with combined (ensemble) algorithms which involves application of multiple algorithms together [1].

Some notable machine learning algorithms include Artificial Neural Network (ANN), which has an adaptive structure and improve its performance by adjusting its parameters during the training process, Support Vector Machine (SVM), which is a vector space-based classification algorithm that finds a linear or nonlinear multidimensional decision boundary that separates a dataset into two distinct parts, K-Means Clustering and K-Nearest Neighbor (KNN) algorithms, which are methods to reveal hidden patterns in data by dividing it into groups containing similar objects [10-13]. Other machine learning algorithms include Logistic Regression (LR), Decision Tree (DT), Extreme Learning Machine (ELM), and XGBoost [1, 4, 14]. Although deep learning is often referred to as a separate category of artificial intelligence, it is often considered as a subcategory of machine learning, as well. Convolutional Neural Networks (CNN) are particularly used in the field of image processing whereas Recurrent Neural Networks (RNN) are mainly utilized for time-series data [15, 16]. Autoencoders are also found to be effective in detecting the relationships in the input data [17]. Some notable studies about the classification of root causes of events can be found in [3, 4, 9, 10, 18-23].

## 3. EVALUATION OF EVENT CLASSIFICATION MODELS

### 3.1. Power Quality Events

The purpose of the event classification is primarily to identify a wide range of root causes of faults and events by recognizing waveforms on the power quality and grid monitoring system. In this study, there are 13 event classes considered, the first 11 of which correspond to transmission system faults: AG, BG, CG, ABG, ACG, BCG, ABCG, AB, AC, BC, ABC faults, line energizing, and line de-energizing. There will also be more additions of root causes of events to be classified by the module in the future such as induction motor starting, transformer energizing, capacitor and reactor switching.

### 3.2 Classification Models and Features

In the feature extraction stage, DWT method has been used. With the DWT method, the three-phase voltage and current signals in each data have been transformed into 5 detailed and 1 approximate coefficient signals using the 5-level db4 mother wavelet. To convert these coefficient signals into scalar magnitudes, statistical features such as mean, standard deviation (SD), root-mean-square (RMS), energy, skewness, kurtosis, Shannon entropy, and maximum bandwidth have been used. These calculated scalar magnitudes are used later as extracted features specific to the signal in the classification stage.

Machine learning algorithms (without deep learning algorithms) have been currently used during the classification stage, yet, as part of future work, deep learning based machine learning algorithms will be employed for classification, as well. The feature set obtained by processing the event data are used during the classification phase, both using the algorithms in the "Classification Learner" module of the MATLAB platform and various machine learning algorithms implemented in the scikit-learn open-source machine learning library[1] implemented with the Python programming language. These algorithms generally include statistical algorithms such as ANN, SVM, DT, and clustering methods. There are 30 different classifiers (including various derivatives of these models) in the "Classification Learner" module of the MATLAB platform, and 8 different classifiers are applied in the Python environment.

### 3.3 Training and Test Datasets

Training and test datasets are created using PSCAD simulation software; however, real events collected from the Turkish electricity transmission grid will be classified in the aforementioned grid monitoring system. The datasets comprise 150 training and 150 testing samples for each class where these samples are produced with the sampling frequency of 4 kHz and each sample includes 3-phase voltage and current data for 0.25 seconds. To ensure data diversity, the time of the fault occurrence, the fault resistance, the location where the fault occurred on the transmission line, the operation time of the circuit breaker, the resistance, inductance and capacitance values of the load were changed as parameters in a wide range.

### 3.4. Evaluation Results

Firstly, 30 machine learning models in MATLAB software are trained and tested on the datasets. The evaluation results are presented in Table 1 where the best and worst performing models and their accuracies are shown in boldface.

*Table 1.* Evaluation Results of Machine Learning Models in MATLAB

| Model | Accuracy | Model | Accuracy |
| --- | --- | --- | --- |
| **Cubic SVM** | **97.5%** | Wide Neural Network (NN) | 92.3% |
| Linear SVM | 97.0% | Trilayered NN | 92.1% |
| Linear Discriminant | 96.7% | Cosine KNN | 92.0% |
| Medium NN | 96.1% | SVM Kernel | 91.0% |
| Medium Gaussian SVM | 96.1% | Gaussian Naive Bayes | 89.8% |
| Coarse Gaussian SVM | 96.0% | Bagged Trees | 92.5% |
| Quadratic SVM | 95.3% | Logistic Regression Kernel | 89.8% |
| Fine KNN | 94.8% | Cubic KNN | 88.5% |
| Narrow NN | 93.9% | Kernel Naive Bayes | 88.2% |
| Boosted Trees | 93.7% | Medium Tree | 87.1% |
| Subspace Discriminant | 93.5% | RUSBoosted Trees | 87.1% |
| Weighted KNN | 93.4% | Fine Tree | 87.0% |
| Bilayered NN | 93.2% | Coarse KNN | 77.1% |
| Subspace KNN | 92.8% | Fine Gaussian SVM | 36.6% |
| Medium KNN | 92.5% | **Coarse Tree** | **29.3%** |

Next, similar tests are performed using the open-source scikit-learn machine learning library written in Python and evaluation results of 8 machine learning models are shown in Table 2.

---

1  https://scikit-learn.org/

*Table 2. Evaluation Results of Machine Learning Models in scikit-learn Library*

| Model | Accuracy | Model | Accuracy |
|---|---|---|---|
| **XGBoost** | **97.0%** | RBF SVC | 92.7% |
| Linear SVC | 96.6% | Decision Tree | 91.9% |
| Logistic Regression | 94.1% | Random Forest | 90.7% |
| KNN | 93.3% | **Gaussian Naive Bayes** | **90.2%** |

As the best performing algorithm, the confusion matrix of the Cubic SVM algorithm is presented in Figure 1.

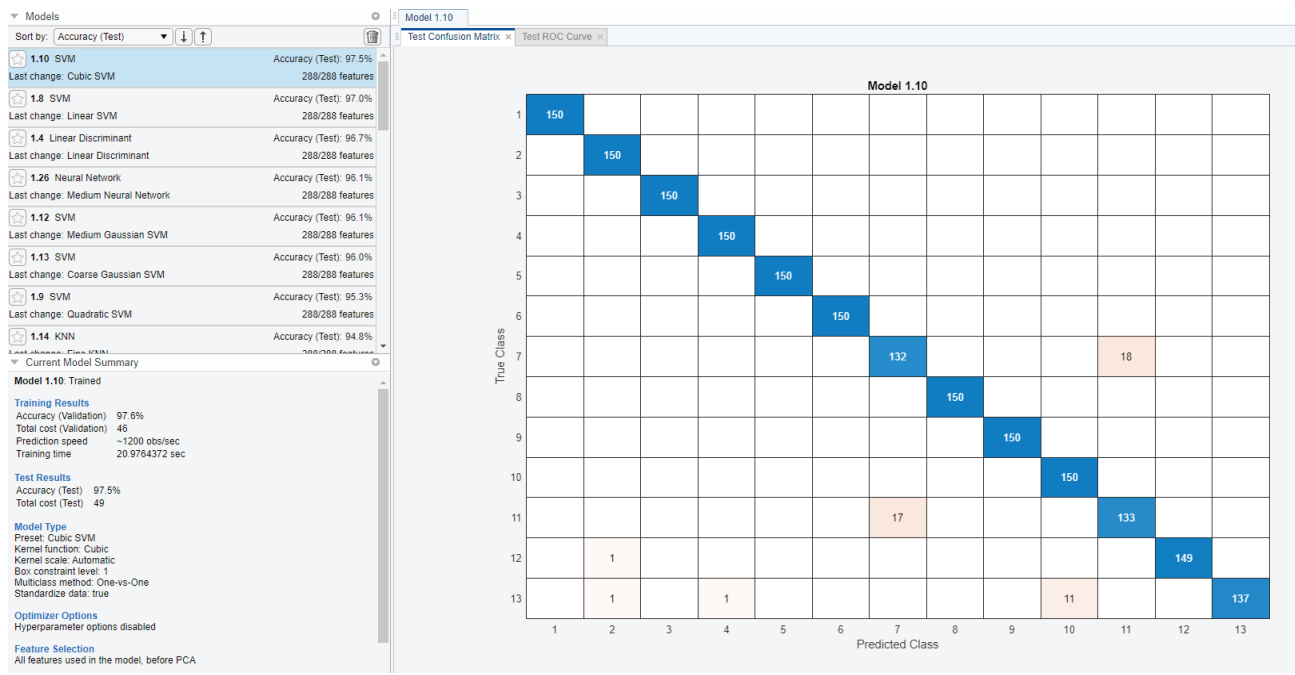

*Figure 1.* Confusion Matrix for the Cubic SVM Algorithm

## 4. DISCUSSION OF THE RESULTS

It can be seen from Table 1 and 2 that the Cubic SVM algorithm, which is a variation of the SVM algorithm, succeeded the classification of 13 different event cause types with a high accuracy of 97.5% in MATLAB whereas the XGBoost algorithm, which is a DT algorithm, achieved this classification with a high accuracy of 97% in the scikit-learn library of Python. As can be observed from the confusion matrix in Figure 1, the algorithm classifies almost all event cause types with high accuracy, but makes a few mistakes when distinguishing between actual ABC and ABCG faults, which stands as the main source of performance degradation for both models.

As part of future work based on the current study:

- Similar experiments will be performed after fine-tuning model parameters.
- Deep learning models will be tested and compared with machine learning models.
- A wider range of (finer granularity) event classes will be considered during classification.
- Other statistical features and feature extraction methods will be applied to the current problem settings.
- The best performing machine learning models will be integrated into the large-scale electric power quality and grid monitoring system (called TEKİS), a preliminary version of which was presented in [24].

## 5. CONCLUSION

In this study, we present the evaluation results of an extensive set of machine learning models for the classification of power quality events with respect to their root causes. The corresponding tests are performed in two distinct platforms: MATLAB tool and scikit-learn open-source machine learning platform. It is observed that the highest performing model in the MATLAB platform is Cubic SVM and the best performing model in scikit-learn library is XGBoost. The best performing models will be integrated into the event classification module of a large-scale nationwide grid monitoring system. Deep learning models and finer granularity event classes will be employed as part of future studies.

## 6. ACKNOWLEDGMENT

**Electric Transmission**

**Oral Presentation:** Classification of Power Quality Events in the Transmission Grid: Comparative Evaluation of Different Machine Learning Models

4th SEERC CONFERENCE ISTANBUL